# Sunspot Records by Antonio Colla just after the Dalton Minimum

V.M.S. Carrasco[1,2,*], C. Bertolin[3], F. Domínguez-Castro,[4,5], L. de Ferri[3], M.C. Gallego[1,2], J.M. Vaquero[2,6]

[1] Departamento de Física, Universidad de Extremadura, 06071 Badajoz, Spain

[2] Instituto Universitario de Investigación del Agua, Cambio Climático y Sostenibilidad (IACYS), Universidad de Extremadura, 06006 Badajoz, Spain

[3] Department of Mechanical and Industrial Engineering, Norwegian University of Science and Technology, Norway

[4] Fundación ARAID, 50018 Zaragoza, Spain

[5] Departamento de Geografía y Ordenación del Territorio, Universidad de Zaragoza, 50009, Zaragoza, Spain

[6] Departamento de Física, Universidad de Extremadura, 06800 Mérida, Spain

* Corresponding author: vmscarrasco@unex.es

**Abstract:** Antonio Colla was a meteorologist and astronomer who made sunspot observations at the Meteorological Observatory of the Parma University (Italy). He carried out his sunspot records from 1830 to 1843, just after the Dalton Minimum. We have recovered 71 observation days for this observer. Unfortunately, many of these records are qualitative and we could only obtain the number of sunspot groups and/or single sunspots from 25 observations. However, we highlight the importance of these records because Colla is not included in the sunspot group database as an observer and, therefore, neither his sunspot observations. According to the number of groups, the sunspot observations made by Colla are similar as several observers of his time. For common observation day, only Stark significantly recorded more groups than Colla. Moreover, we have calculated the sunspot area and positions from Colla's sunspot drawings concluding that both areas and positions recorded by this observer seem unreal. Therefore, Colla's drawings can be interpreted such as sketches including reliable information on the number of groups but the information on sunspot areas and positions should not be used for scientific purposes.

**1. Introduction**

Solar activity can be related to multiple phenomena. For example, the sunspot area and geomagnetic disturbances are manifestations of solar activity (Usoskin, 2017). The most used index in order to study the long-term solar activity is the sunspot number (Vaquero, 2007; Muñoz Jaramillo and Vaquero, 2019). This index is obtained from direct observations of the sunspot number appeared on the solar disc (Hoyt and Schatten, 1998; Clette et al., 2014). The sunspot number has been more or less systematically recorded since 1610, after the invention of the telescope (Arlt and Vaquero, 2020). Thus, this observational set is considered like the world's longest-running experiment (Owens, 2013).

In more than 400 years of telescopic observations, some periods characterized by low solar activity have been recorded. The only grand minimum observed in these four centuries was the Maunder Minimum occurred between 1645 and 1715 (Eddy, 1976; Usoskin et al., 2015; Vaquero et al., 2015). However, other periods of very reduced solar activity, but not considered as grand minima, have also occurred. This is the case of the period corresponding to, approximately, the first third of the 19$^{th}$ century known like Dalton Minimum (McCraken and Beer, 2014). Vaquero et al. (2016), regarding the period 1798–1833, obtained around 60 % of active days *versus* roughly 10 % in the Maunder Minimum. Note that a day is considered active when at least one sunspot was observed on the Sun. We highlight the long sunspot observation series made by Schwabe during 19$^{th}$ century with more than 10000 daily records, including the Dalton Minimum (Arlt et al., 2013; Senthamizh Pavai et al., 2015). Recent works have provided new information about the Dalton Minimum from the recovery of sunspot observations made during this period. For example, Denig and McVaugh (2017) presented the sunspot drawings made by Jonathan Fisher in 1816 and 1817 and Carrasco et al. (2018) analyzed the sunspot observations made by Hallaschka in 1814 and 1816. More recently, Hayakawa et al. (2020) have revised the number of sunspot groups recorded by Derfflinger for the period 1802–1824 and, moreover, they calculated the sunspot positions according to Derfflinger's drawings.

In this work, we analyze the sunspot observations made by Antonio Colla in Parma (Italy) just after the Dalton Minimum. We show some biographical notes about this observer and descriptions about the sunspot records and documentary sources in Section 2. The analysis and discussion of these records are exposed in Section 3. A comparison between the sunspot group number recorded by Colla and Schwabe is discussed in Section 4 and the calculations for sunspot areas and positions from Colla's records are exposed in Section 5. Finally, the main conclusions of this work are presented in Section 6.

**2. Antonio Colla: Biographical Notes and Sunspot Documentary Sources**

2.1. Biographical Notes

Antonio Colla (1806 – 1857) developed his professional career as a meteorologist and astronomer in Parma (Italy) (Cocconcelli, 2009). During the period 1824 – 1828, Colla attended physics classes at Parma University but he interrupted his studies because of economic difficulties. However, he continued studying and carrying out meteorological observations as a self-taught person, staying close to the academic environment. He published his own journal, *Giornale Astronomico ad uso commune*, including observations made the previous year (1829). In this new journal, Colla showed, besides meteorological observations, information on astronomical observations such as moon phases, duration of the day, eclipses and other astronomical phenomena. Colla was recommended by Lucio Bolla, mayor of Parma, to the Queen Maria Luigia and he was hired an employee of the observatory on 6 December 1831 (Lombardini, 1831). However, the official appointment as a meteorological observer was not until 3 November 1834. Antonio Colla was therefore who led the systematic meteorological and astronomical activity in Parma from 1834.

Antonio Colla was appointed Director of the Meteorological Observatory of the University of Parma in 1841 (Rizzi, 1953) and, three years later, he obtained the role of honorary Professor of Meteorology. Because of the new responsibilities and, mainly, his increased interest on seeking comets, Colla definitively stopped sunspot observations in 1840s. He discovered several comets and one of them, C/1847 J1, was later named "Colla" in his honor. Furthermore, Colla was in touch with directors and astronomers of observatories of that time, such as Angelo Secchi, director of the Roman College Observatory (Colombi and Perazzo, 2017). We also highlight his work had recognition from several scientific academies, for example, the Cracow Scientific Society (Poland), the *Académie des Sciences, Arts et Belles-Lettres* of Dijon (France), and the Academy of Sciences of the Institute of Bologna (Italy).

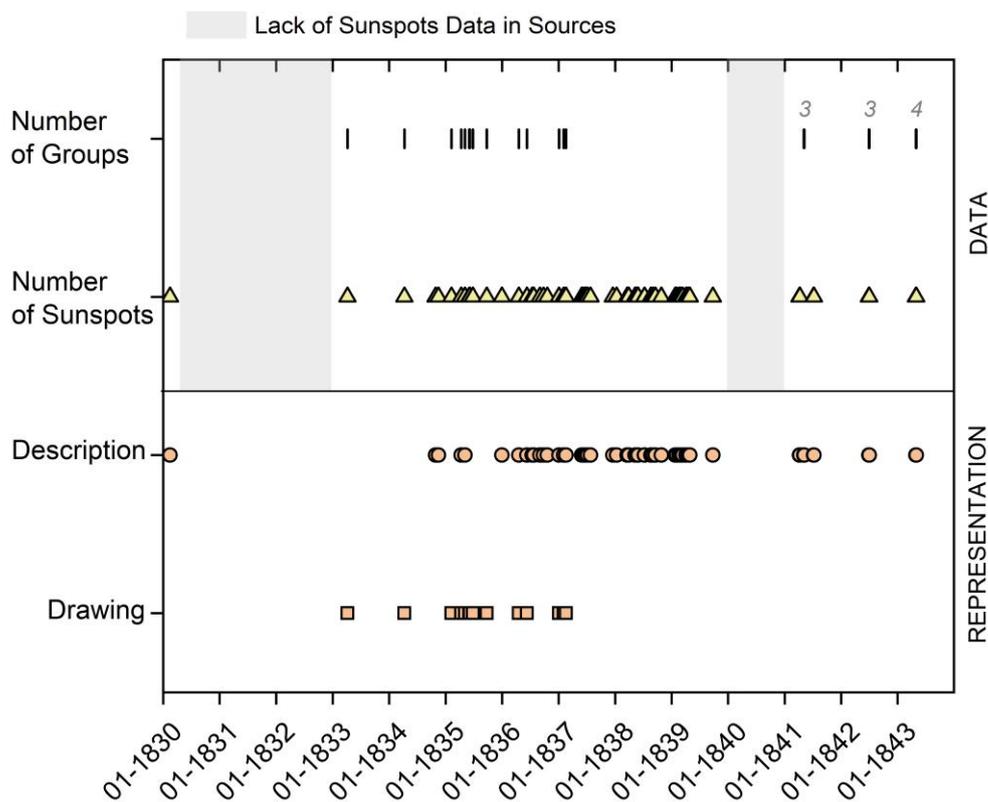

**Figure 1.** Representation of the sunspot record information (circles depict textual descriptions and squares the sunspot drawings) and data availability (triangles represent single sunspots and b sunspot groups). Grey area shows data not available or missed. We specified with numbers "3" and "4" (top right) the number of observations (bars) recovered for those dates because only one bar seems to be represented.

2.2. Sunspot Documentary Sources

We have found sunspot data recorded by Colla in different documentary sources by short descriptions and sunspot drawings (Figure 1). Regarding the first ones, short qualitative information on number of sunspots are provided in the documentary source *Giornale Astronomico* for the year 1830 and from 1834 to 1839. These data were described in the section *Fenomeni notati a Parma nell`anno* (Phenomena noticed in Parma during the year) using frequently adjectives such as *numerose macchie solari* (several sunspots) (Colla, 1833, 1838, 1839, 1840). An example of sunspot data collected in this source corresponding to 11 January 1838 can be seen in Figure 2 (top-left panel).

More information on sunspot descriptions recorded by Colla can be found in *Gazetta di Parma*, i.e. Parma Gazette. An example page of the sunspot records made by Colla in this source for 5 – 7 May 1841 is represented in Figure 2 (bottom panel). The information related to sunspots data was provided in brief articles with title *Varietà astronomiche o meteorologiche* (Astronomical or meteorological varieties) where Colla reported local news but also scientific discovering by other European observers. This kind of news published in *Gazetta di Parma* was also recorded in the meteorological observations made in the Parma Observatory for the year 1841-1843 (Colla, 1842, 1843, 1844). The sunspot information is included in the last column of the table with the title *Note Diverse – Astronomie* (Miscellaneous Notes – Astronomy). Here, it is possible to extract qualitative and sometimes quantitative information on the number of single sunspots and sunspot groups (Figure 2).

The last documentary source where we found sunspot records are the manuscripts included in the meteorological records preserved at the Historical Archive of the Parma University and previously in the library of the meteorological service of the Parma University (Colla 1836, 1837, 1838b). These manuscripts contain the sunspot drawings made by Colla for the period 1835 – 1837. It can be seen the drawings are usually shown after a summary of the meteorological observation by tables or textual descriptions (Figure 2, top right panel). In the drawings, Colla provides the date and hour of the observation. The number of sunspot groups could be obtained for that period.

We have found information on the telescopes of the observatory in the documentary sources. Their descriptions have been found in the inventory of the observatory provided to Colla in 1834 when he was officially appointed as an observer. Three different telescopes are included in that list: i) one achromatic telescope with 17 lines of aperture and 3 feet of focal length, ii) one Gregorian reflector telescope with 2 feet and 4 inches in length and 3 and a half inches in aperture, and iii) a second Gregorian reflector with 4 feet, 1 inch, and 6 lines in length and 7 inches and 2 lines in aperture. We highlight that we have not found explicit information if these values correspond to historical Parma measurements. Note that 1 Parma foot is equal to 0.545 m. Moreover, other four additional tubes are listed but without information about their characteristics. We note Colla employed a helioscope to project sunspots on a sheet but, unfortunately, we do not know exactly what telescope he used in his sunspot observations.

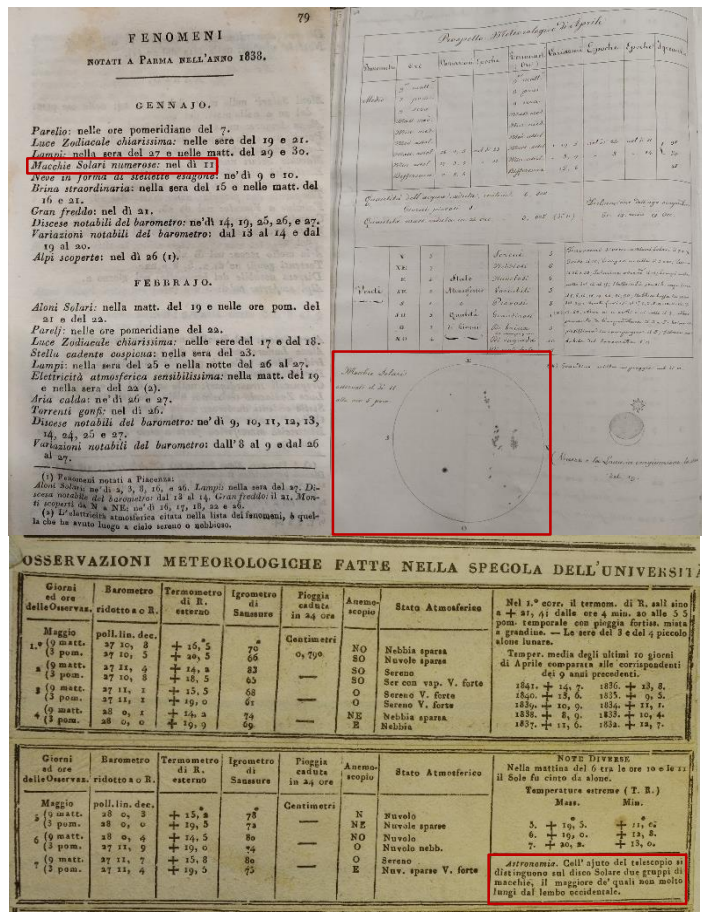

**Figure 2.** Example pages of the documentary sources where Colla's sunspot observations were found: top left panel - *Giornale Astronomico*, top right panel – manuscript records with sunspot drawing, and bottom panel - *Gazetta di Parma*, and. Red squares contain the information relative to sunspots recorded by Colla [Sources: Colla, 1839, 1842, 1837].

## 3. Analysis of the Colla's sunspot observations

Colla carried out sunspot observations at Parma (Italy) from 1830 to 1843. We have recovered 71 observation days made by Colla in that period: 1 in 1830, 2 in 1834, 10 in 1835, 7 in 1836, 10 in 1837, 14 in 1838, 11 in 1839, 9 in 1841, 3 in 1842, and 4 in 1843. Colla published sunspot drawings in 14 of these observation days: 9 in 1835 (7 February, 6, 9, and 12 April, 6 May, 3, 6, and 26 June and 24 September), 2 in 1836 (18 April and 10 June), and 3 in 1837 (3 January and 2 and 17 February). Unfortunately, Colla did not provide quantitative information on the sunspot or group number in 46 observation days. For example, Colla (1838a) pointed out that numerous sunspots were observed on 28 October and 16 November 1834 without specifying the number of single sunspots or groups. Colla recorded in a general way the number of groups for three periods: 5 – 7

May 1841, 29 June – 1 July 1842, and 29 April – 2 May 1843. As an example, Colla (1842) recorded for 5, 6, and 7 May 1841 the following note: "Astronomy. With the help of the telescope, two groups of sunspots were observed, the largest not far from the western edge". Moreover, on 14 February 1830, Colla provided information on single sunspots, but not on the group number: "… and, on 14 February 1830, three very large sunspots were on the solar disc" (Colla, 1833). In summary: i) we found information both the number of groups and single sunspots in 14 days: 7 February, 6, 9, and 12 April, 6 May, 3, 6, and 26 June and 24 September 1835, 18 April and 10 June 1836, and 3 January and 2 and 17 February; ii) we obtained information only of the number of groups in 10 days: 5 – 7 May 1841, 29 June – 1 July 1842, and 29 April – 2 May 1843; and iii) we only extracted only the number of single sunspots in 1 day: 14 February 1830.

Colla carried out his sunspot observations during Solar Cycle 8 except one observation around the maximum of Solar Cycle 7, on 14 February 1830, when he reported three single sunspots. According to the smoothed sunspot number version 2 (http://sidc.be/silso), Solar Cycle 8 started in November 1833 and finished in July 1843 (minimum of Solar Cycle 9). The maximum amplitude of Solar Cycle 8 lies in March 1837. This is the solar cycle of the 19th century with the maximum amplitude. The observational coverage for Solar Cycle 8 according to the current group database (Vaquero et al., 2016) is 72.8 % and the most active observers in this cycle were Schwabe (2386 observations days), Schmidt (458) and Stark (332). Figure 3 depicts the daily number of groups recorded by Colla (red color) and that for all the observers included in the current group database (black color). The single sunspot and sunspot group counting carried out in this work is publicly available on the website of the Historical Archive of Sunspot Observations (http://haso.unex.es). We note that, in Colla's drawings, there are some cases in which it is difficult to identify sunspots from other kind of phenomena. For example, our hypothesis is that some of the dot sets recorded by Colla on 10 June 1836 (Figure 4) such as, for example, the long line of very small dots next to the western limb, are facular regions and not sunspots. Most of the records by Colla were made during the rise phase and maximum of that solar cycle. His last observations were in the last part of the decline phase and around the minimum of Solar Cycle 9. The only sunspot records carried out in a different solar cycle was the aforementioned observation corresponding to 14 February 1830 (Solar Cycle 7). The remaining records recorded between 1837 and 1841 do not include quantitative information. The maximum daily number of sunspot groups recorded by Colla was 10 (18 April and 10 June 1836) while the minimum value was 1 around the minima of Solar Cycle 8 and 9. Therefore, Colla did not provide information on the spotless days.

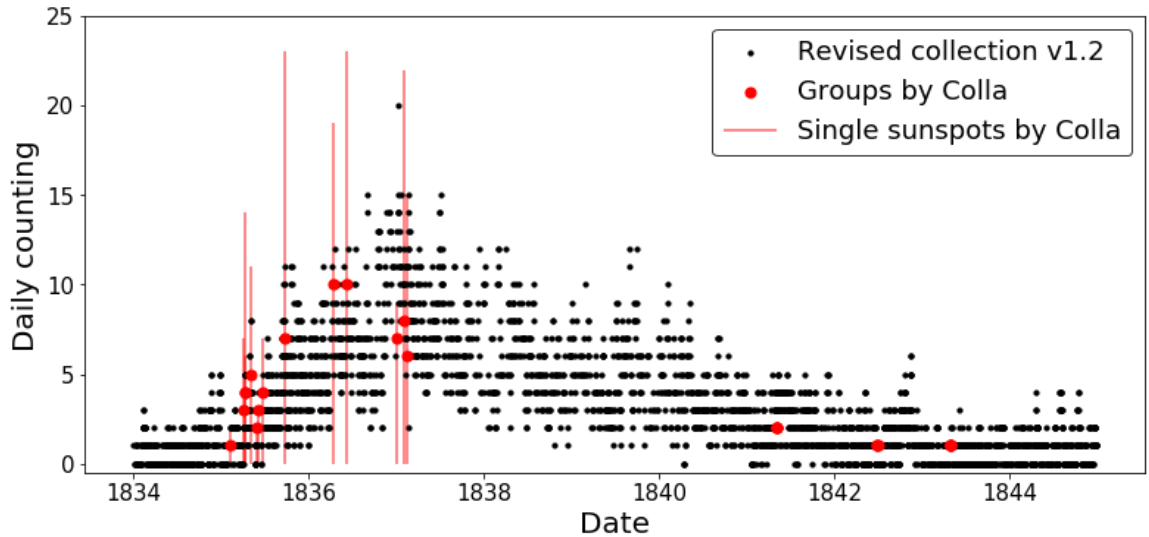

**Figure 3.** Daily number of sunspot groups recorded by Colla (red dots) and those included in the current group database for the period 1834 – 1844 (black dots). Vertical red lines represent the daily number of single sunspots recorded by Colla in his drawings.

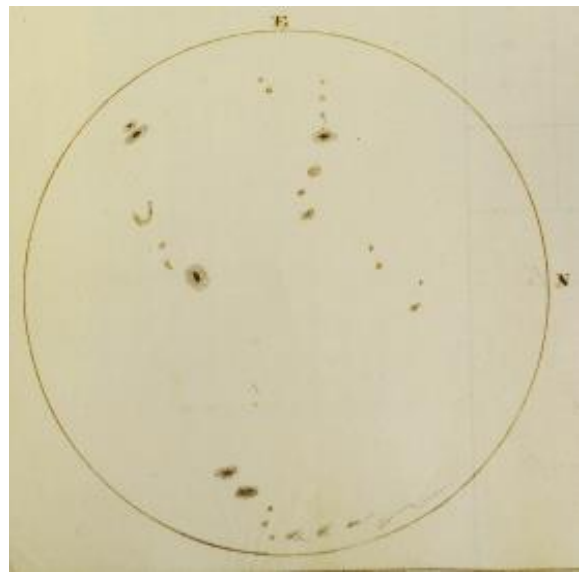

**Figure 4.** Sunspot drawing made by Colla on 10 June 1836 [Source: Colla, 1837].

We acknowledge the number of sunspot records made by Colla with quantitative information including group number, 24 in total, is not a great amount. However, they are important because this observer is not included in the current group database (Vaquero et al., 2016). Thus, this new information can be used to fill gaps in the database and compare them with sunspot observations made by other observers of that time. The sunspot observation made by Colla on 3 January 1837 can fill a gap in the group database because no observation is available for that date. Thus, the observational coverage for that year would increase from 217 to 218 observation days. We note that there is no information in the current group database in 17 of the 46 observation days when Colla provided qualitative information and therefore we can know from Colla's records they were active days.

**Table 1.** Number of common observation days between Colla and other observers and comparison of the daily average of the group number from their sunspot observations.

| Observers | Common observation dates | Average of the daily group number (Colla in bold) |
|---|---|---|
| Schwabe/**Colla** | 21 | 3.8/**3.4** |
| Schmidt/**Colla** | 7 | 1/**1** |
| Kunitomo/**Colla** | 6 | 3.2/**4** |
| Stark/**Colla** | 5 | 6.8/**5.4** |
| Hussey/**Colla** | 4 | 5/**4.5** |
| Herschel/**Colla** | 1 | 9/**8** |

Several astronomers made sunspot observations in the same days in which Colla provided quantitative information on the group number (Table 1): Schwabe (21 coincident observation days), Schmidt (7), Kunitomo (6), Stark (5), Hussey (4), and Herschel (1). We can see that Colla generally recorded more groups than Kunitomo because the average of the daily group number recorded by Colla in the common observations days as Kunitomo was 4 instead of 3.2 for Kunitomo. Schmidt, Hussey and Herschel recorded a similar group number as Colla. The daily average of groups recorded by Colla and Schmidt in the same observation days was 1 in both cases. Regarding the same observation days of Colla and Hussey, Colla recorded in average 4.5 groups per day while Hussey 5. In the only coincident observation day considering the sunspot records made by Herschel and Colla (2 February 1837), Herschel observed 9 groups while Colla recorded 8 groups. Moreover, Colla recorded in average less groups than Stark. The daily average of groups recorded by both observers in the same observation days was 5.4 and 6.8, respectively. The comparison between the sunspot records made by Colla and Schwabe is presented below.

## 4. A comparison with Schwabe's sunspot records

Samuel Heinrich Schwabe made 12414 sunspot observations between 1825 and 1867 recording 8486 sunspot drawings in his observation days (Arlt et al., 2013; Senthamizh Pavai et al., 2015). Taking into account the number of sunspot observations, Schwabe is one of the main sunspot observers of the telescopic era and the observer with the largest number of sunspot records of his time. His experience in the sunspot observation allowed him to be the first observer to estimate the period of solar cycle in roughly 10 years. Regarding the works about the Schwabe's sunspot records, we highlight the great effort made by Arlt et al. (2013) to calculate the sunspot areas and positions of the sunspot recorded by Schwabe in his drawings.

**Table 2.** Number of sunspot groups recorded by Colla and Schwabe in the 21 common observation days.

| DATE | COLLA | SCHWABE | DATE | COLLA | SCHWABE |
|---|---|---|---|---|---|
| 6 Apr 1835 | 3 | 3 | 5 May 1841 | 2 | 4 |
| 9 Apr 1835 | 4 | 5 | 6 May 1841 | 2 | 3 |
| 12 Apr 1835 | 4 | 4 | 7 May 1841 | 2 | 3 |
| 6 May 1835 | 5 | 5 | 29 Jun 1842 | 1 | 2 |
| 3 Jun 1835 | 2 | 3 | 30 Jun 1842 | 1 | 1 |
| 6 Jun 1835 | 3 | 3 | 1 Jul 1842 | 1 | 1 |
| 26 Jun 1835 | 4 | 4 | 29 Apr 1843 | 1 | 1 |
| 24 Sep 1835 | 7 | 7 | 30 Apr 1843 | 1 | 1 |
| 18 Apr 1836 | 10 | 10 | 1 May 1843 | 1 | 1 |
| 10 Jun 1836 | 10 | 10 | 2 May 1843 | 1 | 2 |
| 17 Feb 1837 | 6 | 7 | | | |

Schwabe was a much more active sunspot observer than Colla. For the observation period of Colla (14 February 1830 – 2 May 1843), Schwabe made 3247 sunspot observations and Colla made 71 (including his non-quantitative sunspot records). Taking into account days when Colla provided quantitative information on the group number, the number of observation days in which Colla and Schwabe recorded sunspot records simultaneously were 21 (Table 2). Figure 5 represents the daily group number recorded by these two observers in days when both observers recorded observations. Although the daily group number recorded by Schwabe was always greater or equal to that recorded by Colla, we can see that there is a general agreement. Thus, the daily average of the group number recorded by Colla and Schwabe in those 21 observation days was 3.4 and 3.8, respectively. Schwabe recorded more group numbers than Colla in 8 observations days while it was the same in the remaining 13 observations. The difference in the group number recorded by Schwabe and Colla when Schwabe recorded more groups than Colla was only one, except on 5 May 1841 when Schwabe recorded four groups and Colla only two. The expression for the best linear fit regarding the group number recorded by Colla and Schwabe is: $G_S = (0.95 \pm 0.05)\ G_C + (0.61 \pm 0.21)$, $r = 0.977$. $p$-value $< 0.001$, where $G_S$ and $G_C$ are the daily group number recorded by Schwabe and Colla, respectively. If we set the y-intercept of the line equal to zero, we obtain a linear calibration factor next to 1: $k_C = 1.06 \pm 0.04$, $r = 0.989$, $p$-value $< 0.001$.

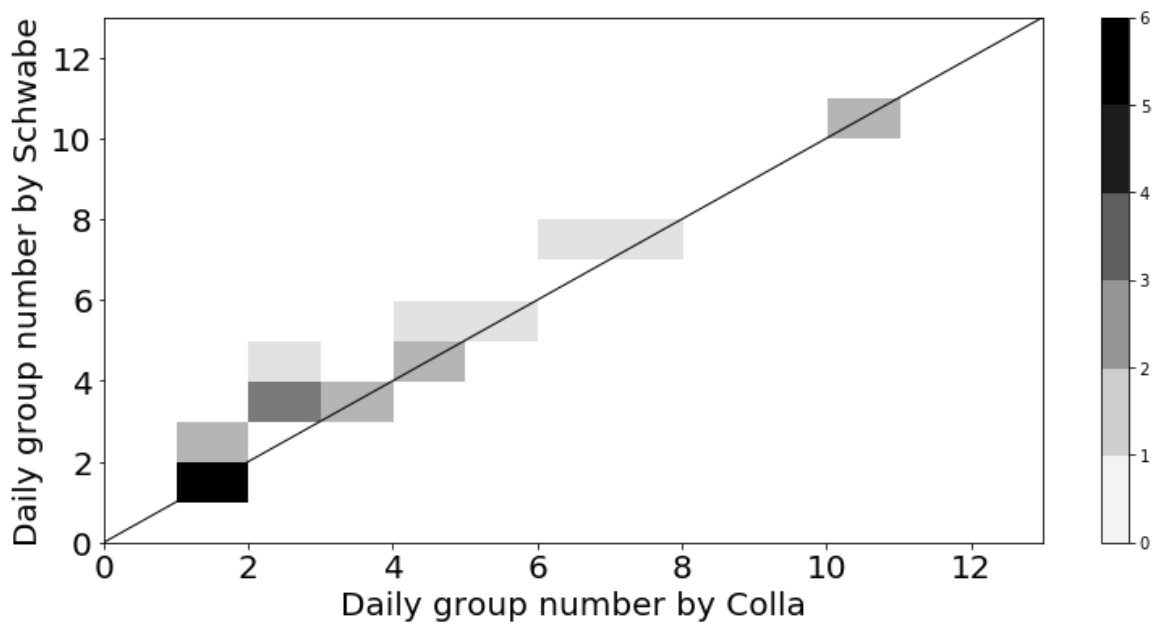

**Figure 5.** Daily number of sunspot groups recorded by Colla and Schwabe. Only coincident observation days have been taken into account in each case. Different colors

represent the number of days in which a specified combination of the number of groups recorded by the two observers is given. Diagonal line depicts the line of slope equal to 1.

## 5. Areas and sunspot positions

We have calculated the sizes of the sunspots and heliographic latitudes from the 14 sunspot drawings recorded by Colla. In order to carry out this task, we have employed *Soonspot*, software created by Galaviz et al. (2020) to determine sunspot areas and positions. First, the sunspot areas recorded by Colla seem exaggerated. Figure 6 (left panel) depicts the sunspot drawing made by Colla on 9 April 1835. We highlight that the area corresponding to the sunspot next to the western limb ("O" in the drawing) is around 2300 millionths of solar hemisphere (msh), a value quite big considering one sunspot only. Moreover, the areas of the other two largest sunspots recorded in that drawing are around 700 and 1000 msh and even the smallest sunspots have areas greater than 50 msh. Note that penumbras typically appear when the sunspot area is greater than 50 msh (Hathaway, 2015). As an example, the group composed by the southernmost four sunspots in the drawing (Figure 6, left panel) have areas greater than 100 msh and however they seem simple pores according to the drawing. We highlight that these values of sunspot areas are recurrent in the remaining drawings. Furthermore, we have not calibrated the sunspot areas recorded by Colla because the number of cases is too low to make a statistical comparison with meaningful results between sunspot area distributions (Baranyi *et al*., 2001).

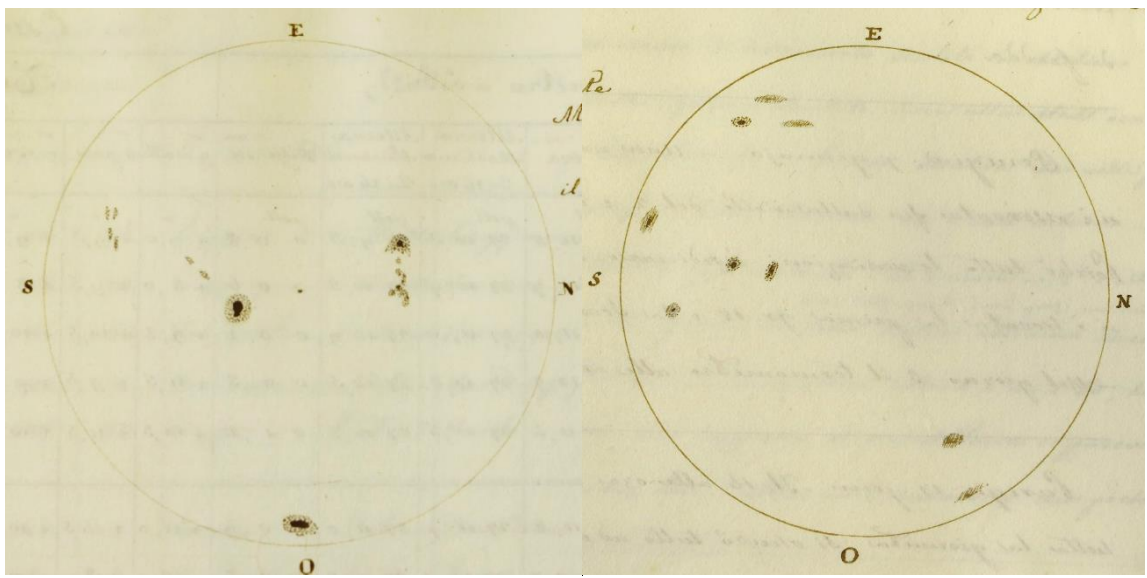

**Figure 6.** Sunspot drawing made by Colla on 9 April 1835 (left panel) and 3 January 1837 (right panel). Note that the small central black dot in the left drawing represents the center of the solar disk [Source: Colla, 1836, 1838b].

We have also calculated the sunspot positions from Colla's drawings. Figure 7 depicts the heliographic latitudes obtained from the sunspot drawings made by Colla and those calculated by Arlt et al. (2013) for Schwabe. We highlight that the values obtained seem unreal. An example can be seen in Figure 6 (right panel) where the sunspot drawing made by Colla on 3 January 1837 is depicted. We note that sunspots are represented with respect to the celestial axes and not to the solar axes. Most of sunspots were recorded within the range from 30º to -30º in latitude for that day. These values are frequently observed during the maxima of solar cycles and we note the observation was made in 1837, around the maximum of Solar Cycle 8. However, we obtained latitudes of -55º and -65º for the two southernmost sunspots. Therefore, these values do not seem real. Thus, as in the case of the sunspot areas, we should take caution with the sunspot positions recorded by Colla in his sunspot drawings. Because of the problems found in the determination of sunspot areas and positions from Colla's drawings, we can interpret these drawings such as sketches from the original observations where we can know the number of sunspots and groups recorded by Colla. However, these drawings do not include accurate information about sunspot areas and positions to be considered scientific usable data.

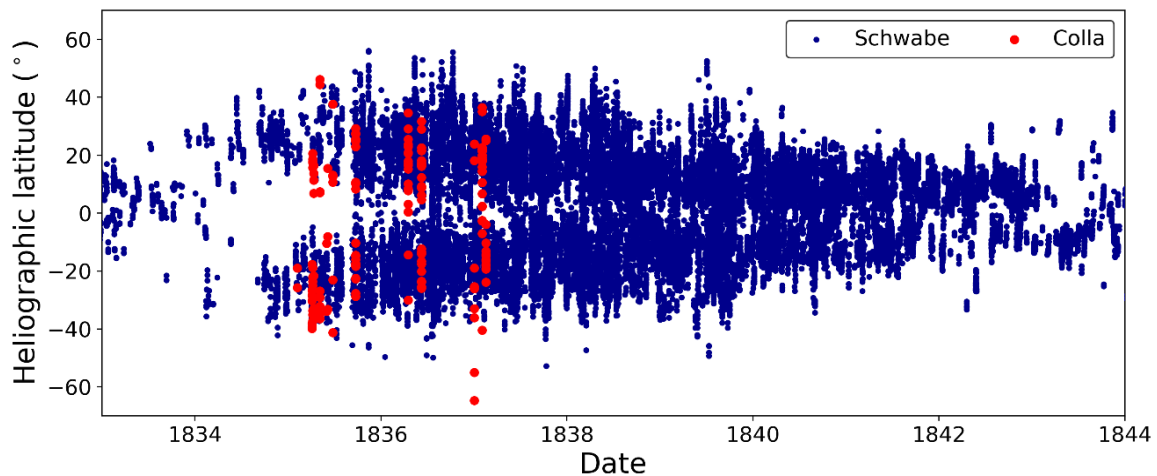

**Figure 7.** Sunspot positions calculated from Colla's sunspot drawings (red color) and those obtained from Schwabe's records (blue color) by Arlt et al. (2013).

## 6. Conclusions

Antonio Colla made sunspot observations in Parma (Italy) for the period 1830 – 1843. We found his sunspot records in three documentary sources. Two of them were by short textual reports: i) *Giornale Astronomico* for the year 1830 and from 1834 to 1839 and ii) *Gazetta di Parma* from 1841 to 1843. The other documentary source are manuscripts where Colla recorded 14 sunspot drawings for the period 1835 – 1837.

Many of the sunspot records made by Colla include qualitative information and, therefore, we could only obtain the number of sunspot groups and single sunspots from 25 observation days (14 days with information both the number of groups and single sunspots, 10 days with information only for groups, and 1 day with information only with the number of single sunspots). The single sunspot and sunspot group counting presented in this work for Colla's sunspot records is publicly available on the website: haso.unex.es. We note Colla only reported active days (days with at least one sunspot on the solar disc) and the maximum number of groups observed by Colla was 10 recorded on 18 April and 10 June 1836. Moreover, one record made by Colla on 3 January 1837 can fill a gap in the group database because no observation is available for that date. Regarding the observation with qualitative information, there is no information in the current group database for 14 days recorded by Colla as active days. We have compared Colla's sunspot observations with the sunspot records made by other observers of that time. According to the average of the group number for the common observations days, we can see that: i) Colla recorded more groups than Kunitomo, ii) a similar number as Schmidt, Hussey, and Herschel, and iii) lower number than Stark. In particular, we have also compared the sunspot observations made by Colla and Schwabe, the most important sunspot observer of that time. Although the average of the daily group number is similar according to the two observers for coincident observations days (3.4 according to Colla and 3.8 to Schwabe), Schwabe always recorded a number of groups greater or equal than Colla. Furthermore, we have calculated the sunspot area and positions from Colla's sunspot drawings. We have obtained areas even equal to 2300 millionths of solar hemisphere for one sunspot, which seems exaggerated. In addition, we have obtained heliographic latitudes up to 65º, significantly different to sunspot positions recorded by Schwabe. We conclude that both sunspot areas and positions recorded by Colla seem unreal and they should not be used for scientific purposes. Thus, Colla's sunspot drawings should be interpreted such as sketches from the original observations with reliable information on the number of groups recorded by Colla but not on sunspot areas and positions.

This work is part of the global effort to provide a sunspot group database with a better observational coverage and good quality observations (Muñoz-Jaramillo and Vaquero, 2019). It helps us to reconstruct as reliable as possible the sunspot number index. However, we still have a long way ahead and this work of recovery and analysis of sunspot records, in particular for the earliest observations of the telescopic era, must be continued.

**Acknowledgments**

This work was partly funded by FEDER-Junta de Extremadura (Research Group Grant GR18097 and project IB16127) and from the Ministerio de Economía y Competitividad of the Spanish Government (CGL2017-87917-P). The authors have benefited from the participation in the ISSI workshops led by M.J. Owens and F. Clette on the calibration of the sunspot number. The authors are grateful to Mr. P Fantini of the Meteorological Service and Dr. Maria Grazia Perazzo of the Historical Archive of the Parma University for the availability in consulting the original sources and to Dr. Ilaria Alfieri, Department of Chemistry, Life and Environmental Sustainability Sciences, University of Parma for having retrieved high resolution images of the sunspots drawings.

**Disclosure of Potential Conflicts of Interest and Ethical Statement**

The authors declare that they have no conflicts of interest. All authors contributed to the study conception and design. The first draft of the manuscript was compiled by V.M.S. Carrasco and all authors read and approved the final manuscript.

**Appendix 1: Historical Sources**

Archivio di Stato di Parma. 1828a. Presidenza dell`Interno, busta 48. Nota statistico personale del Signor Colla Antonio, del 22 Maggio 1828.

Archivio di Stato di Parma. 1828b. Presidenza dell`Interno, busta 48. Documento del 16 Maggio 1828.

Archivio di Stato di Parma. 1828c. Presidenza dell`Interno, busta 48. Protocollo del 20 Giugno 1828.

Archivio Storico dell`Università di Parma. 1834. Divisione unica, busta 31, fascicolo 4, carta 54. Manuscript.